\documentclass{PoS}

\title{Hearing the echoes of dark matter and new physics}

\ShortTitle{Hearing the echoes of dark matter and new physics}

\author{\speaker{Fa Peng Huang}\\
        Center for Theoretical Physics of the Universe,
Institute for Basic Science (IBS), Daejeon 34126, Korea\\
        E-mail: \email{huangfp@ibs.re.kr}}

\abstract{Motivated by the absence of dark matter signal in dark matter direct detection 
experiments and new physics signal at LHC, we study how to hear the echoes 
of the new physics, especially the dark matter and baryogenesis by new approaches,  
such as the gravitational wave experiments.}

\FullConference{ICHEP2018, XXXIX International Conference on High Energy Physics\\
		4-11 July 2018\\
Seoul, Korea}

\begin{document}

\section{Introduction}

After the discovery of gravitational wave (GW) by aLIGO,
GW becomes a new approach to explore  the new physics (NP) beyond the standard model and the fundamental problems in particle cosmology.
Especially, obvious shortcomings in our understanding of the dark matter (DM) and the baryon asymmetry of the universe (BAU), and no evidence of NP at LHC may 
give us a hint that  a new approach is needed to explore DM and BAU.
GW may be used to hear the echoes of DM, baryogenesis, NP models, symmetry breaking patterns of the early universe. 
We study how to detect DM and NP by GW.
Generally, strong first-order phase transition (SFOPT)  may be induced in many other NP models, which may produce detectable GW
through bubble collisions, turbulence and sound wave mechanisms.
There are two typical examples.
One is the SFOPT triggered by the DM,
the other is the SFOPT  in the baryogenesis mechanism.

\section{Probing dark matter and baryogenesis relaxed  by phase transition}
\begin{figure}[h]
\begin{minipage}[t]{0.3\linewidth}
\centering
 \includegraphics[width=1.0\linewidth]{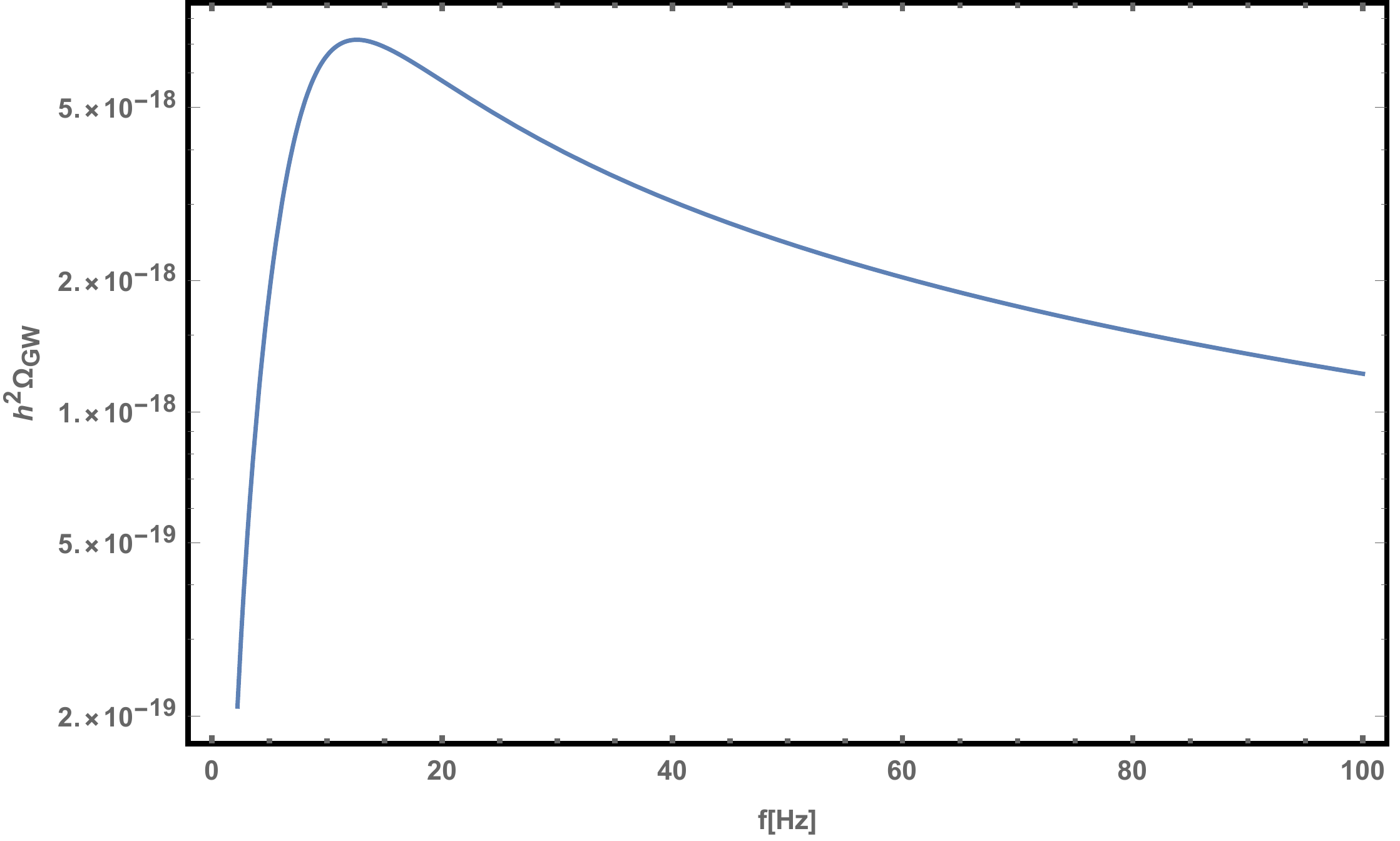}\\
\end{minipage}
\hfill
\begin{minipage}[t]{0.3\linewidth}
\centering
 \includegraphics[width=1.0\linewidth]{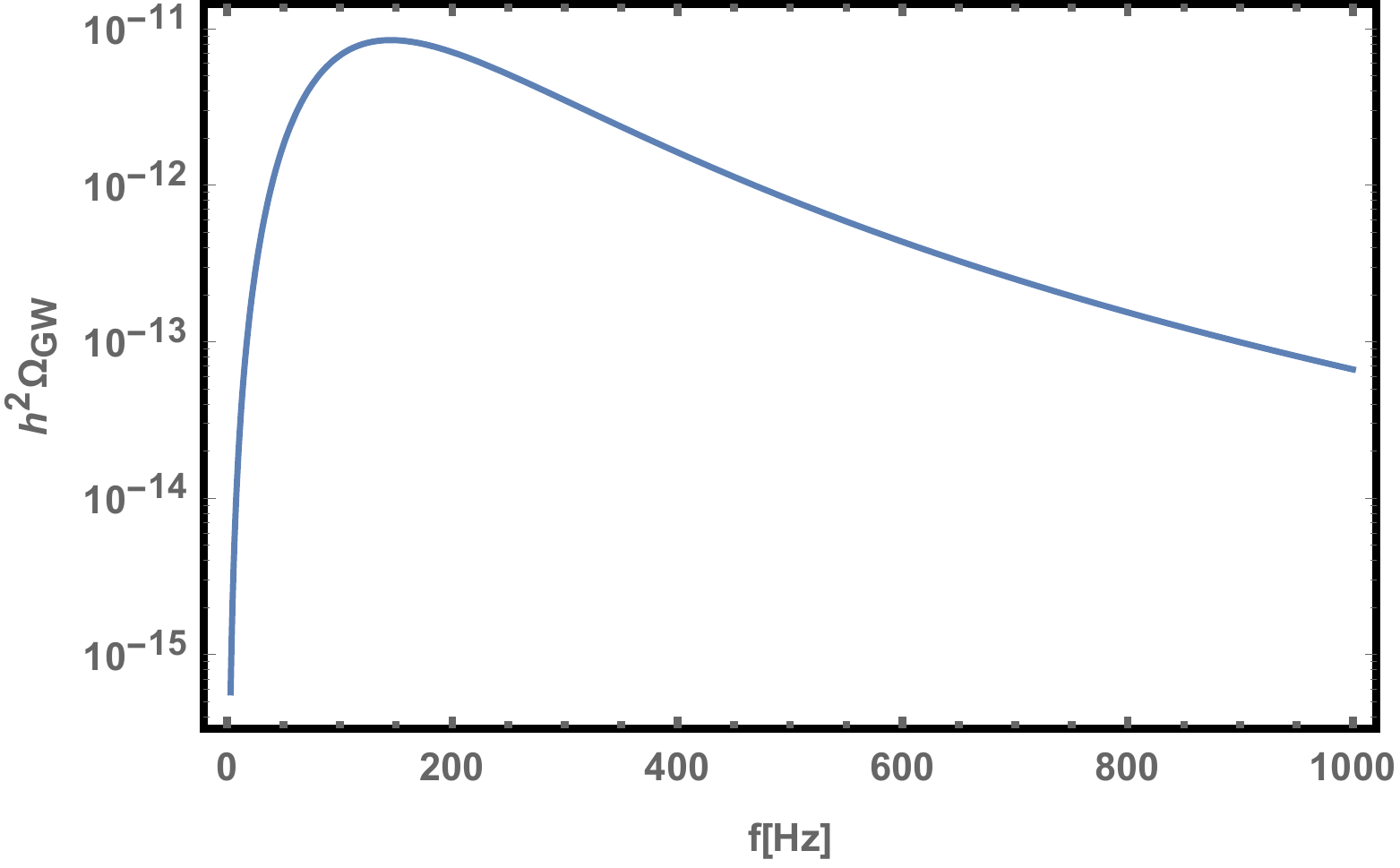}\\
\end{minipage}
\hfill
\begin{minipage}[t]{0.3\linewidth}
\centering
 \includegraphics[width=1.0\linewidth]{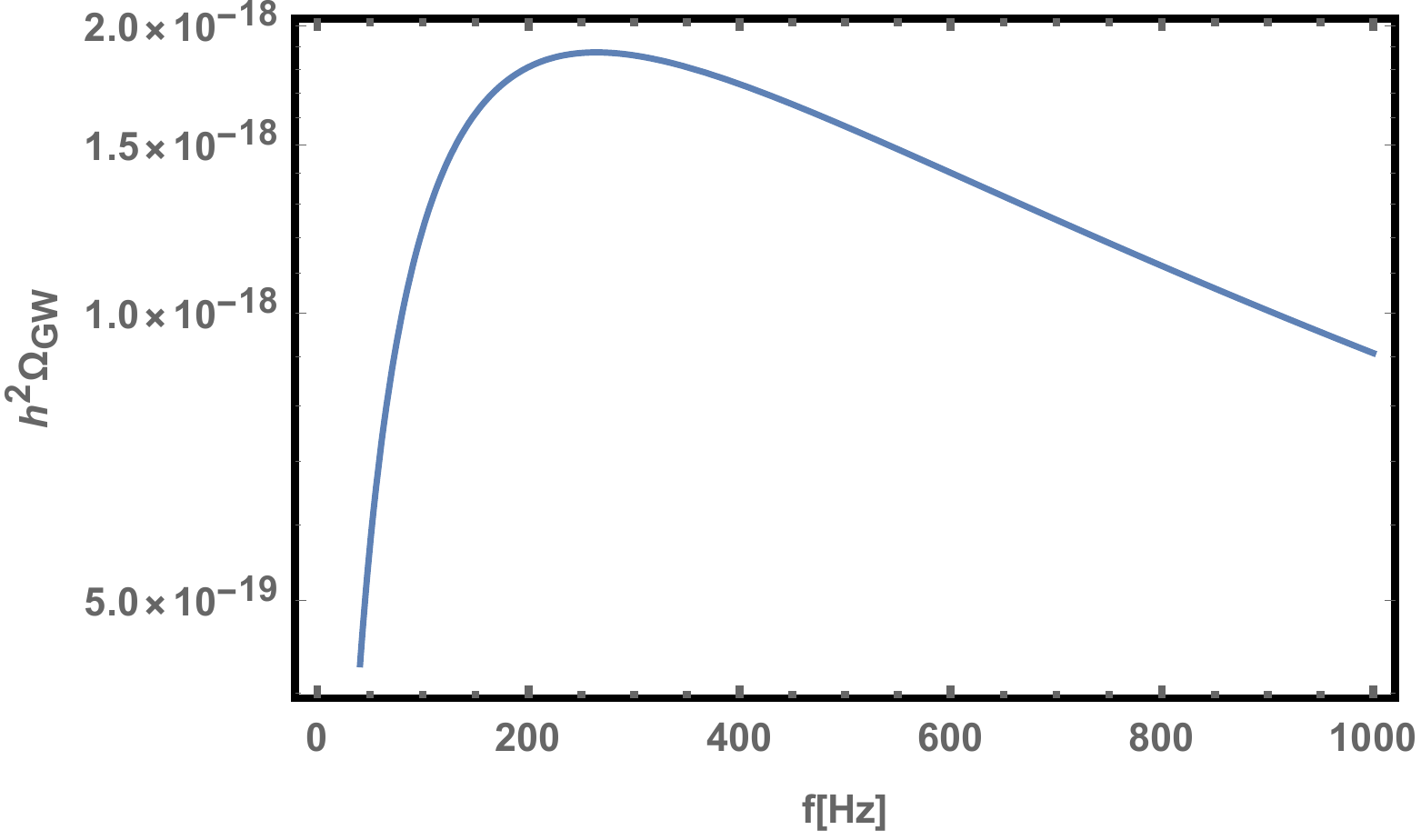}\\
\end{minipage}
\caption{The predicted GW spectrum for benchmark I with $v_b=0.3$.  From left to right, the line represents the GW 
spectrum from bubble collision, sound waves and turbulence, respectively.
}\label{gw_b1}
\end{figure}

In usual asymmetry DM models to explain the DM and BAU, there are usually 
strong constraints on the DM mass and the reheating temperature.  The DM should be about 5 GeV, and the
reheating temperature should be fine-tuned to avoid washout process.
By assuming SFOPT with Q-ball production, the constraints can be 
significantly relaxed~\cite{Huang:2017kzu}. We start with the following Lagrangian:
\begin{eqnarray}\label{effsum}
	\mathcal{L} &=& \frac{1}{2}(\partial_\mu S)^2- U(S)+ (\partial_\mu \chi)^{*}(\partial_\mu \chi)-k_1^2 S^2 \chi^{*} \chi -\sum_{i}\frac{h_i^2}{2} S^2 \phi_i^2\nonumber \\
&+&\sum_{i}\frac{1}{2}(\partial_\mu \phi_i)^2	-\!\!\sum_{a=1,2}\!\!\frac{\lambda^{ijk}_a}{\Lambda^2}\bar{X}_a P_R D_i \bar{U}^{C}_j P_R D_{k} +\frac{\zeta_a}{\Lambda} \bar{X}_a Y^C \chi \chi^* + {\rm H.c.}
\end{eqnarray}
with $U(S)=\lambda_S (S^2-\sigma^2 )^2/4 $.
In the every early universe, the scalar field $S$ has no vacuum expectation  value (VEV) and 
BAU can be produced by heavy particle decay with the interference effects between 
two-loop and tree-level diagram.
After the BAU is produced, the scalar field $S$ acquires a VEV and SFOPT occurs.
With the expansion of the bubbles, the  would-be DM particles $\chi$  are packeted into the
Q-balls which is guaranteed by the global $U(1)$ symmetry. Q-balls are non-topological solitons.
To explain the observed DM and BAU, the final condition can be  obtained as 
\begin{equation}\label{bdm}
	\rho_{DM}^4  v_b^{3/4}  =73.5 (2 \eta_B s_0)^3 \lambda_S \sigma^4 \Gamma^{3/4}  \,\,\, \, .
\end{equation}
The  typical benchmark points are shown  in Tab.~\ref{benp}.
\begin{table}[h]
\begin{center}
\begin{tabular}{|c|c|c|c|c|c|}
   \hline
   benchmark sets   &$\lambda_S$& $e$    &$c$&$T_c$ [TeV]& $\frac{\sigma}{T_C}$     \\
   \hline
                 I  & 0.008     & 0.754  & 1 & 15.9      &                  5             \\
   \hline
                 II & 0.0016    & 0.151  & 1 & 6.6       &                  5           \\
   \hline
 \end{tabular}
 \end{center}
 \caption{The benchmark sets to satisfy the observed DM density and BAU with $v_b=0.3$. }\label{benp}
\end{table}
And the corresponding GW signals are shown in Fig.~\ref{gw_b1}. This scenario can also have abundant collider phenomenology at LHC.
We discuss the collider constraints and predictions at QCD next-leading-order level in Ref.~\cite{Huang:2017kzu}.

\section{Exploring the blind spots of dark matter models}
\begin{figure}
\begin{center}
\includegraphics[scale=0.35]{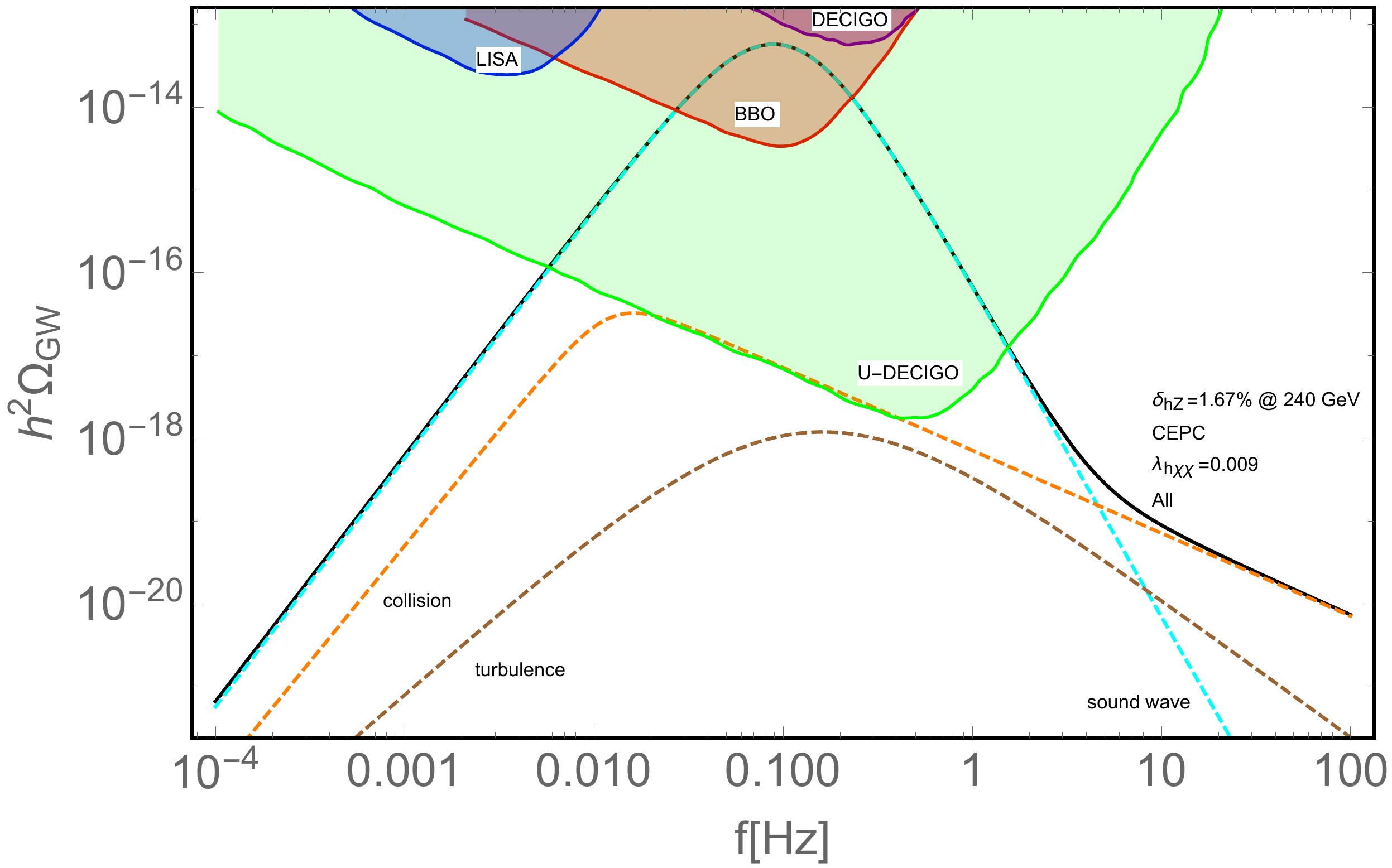}
\caption{The phase transition GW spectra $h^2\Omega_{\rm GW}$ for the benchmark set in the inert doublet model.}
\label{idm2}
\end{center}
\end{figure}

Motivated by the absence of DM signal in DM direct detection  (such as the LUX, PandaX-II, XENON1T), a generic classes of scalar DM models have been pushed to the blind spots where dark matter-Higgs coupling is very small.
We use the complementary searches via phase transition GW and the future lepton collider signatures to un-blind the blind DM spots.
Taking the inert doublet model as an example, we show that GW and CEPC can help to explore the blind spot for this DM model since
the SFOPT can still be induced despite the small coupling between the DM matter and the Higgs boson~\cite{Huang:2017rzf}.
Considering the DM constraints, the collider constraints and the condition for SFOPT, we show 
one set of benchmark points $\lambda_3=2.84726$, $\lambda_4=\lambda_5=-1.41463$ and $M_D=59.6\rm~GeV$ here, where
the corresponding DM mass is 64 GeV, the pseudo scalar mass and the charged scalar mass are both 299.6 GeV,
$\lambda_{h\chi\chi} =\lambda_{345}/2=0.009$. The corresponding GW signals and collider signals at CEPC are shown
in Fig.~\ref{idm2}. We also study the GW signals  for mixed  singlet-doublet model and mixed singlet-triplet model in
Ref.~\cite{Huang:2017rzf}.

\begin{figure}
\begin{center}
\includegraphics[scale=0.3]{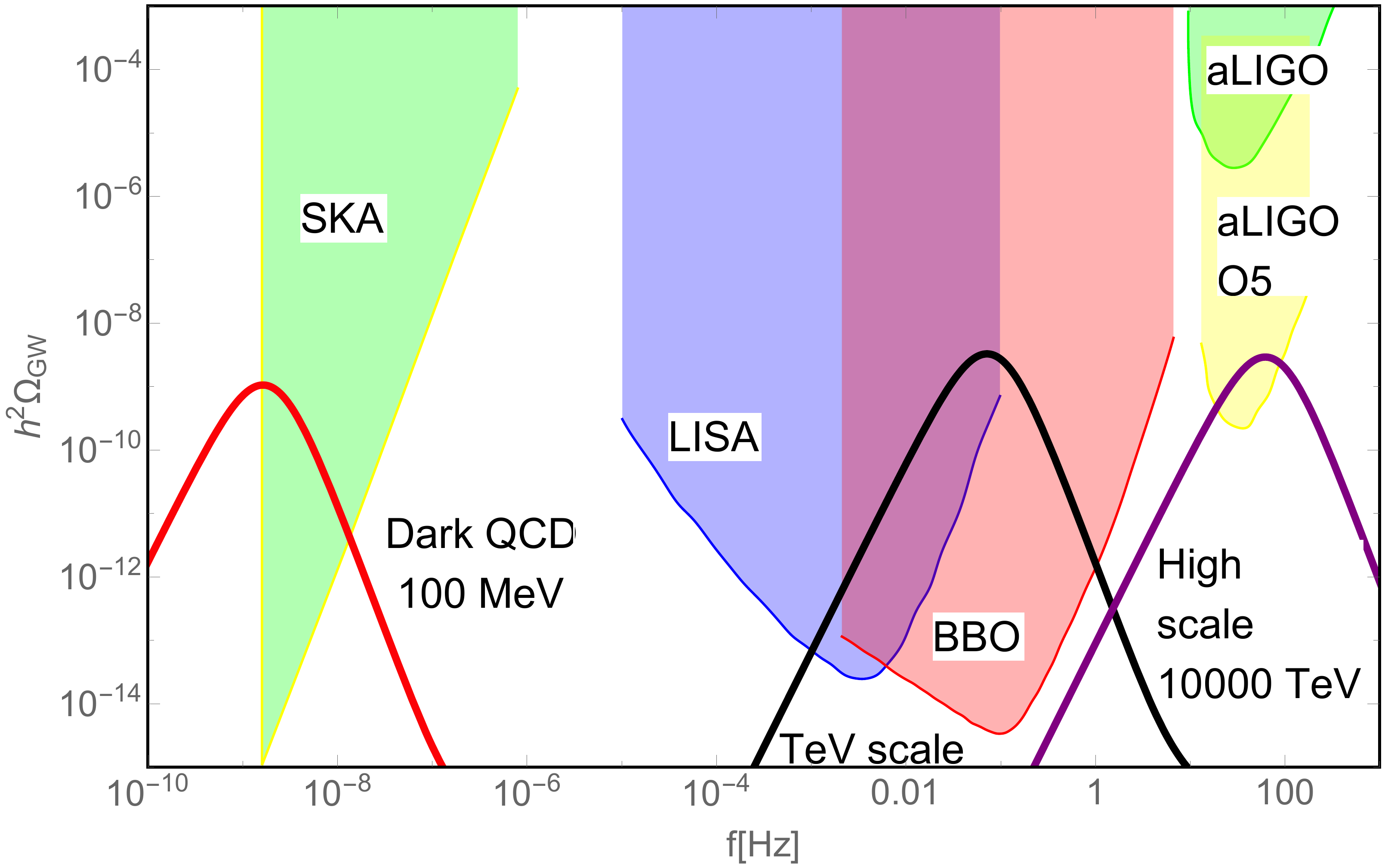}
\caption{Schematic phase transition GW spectra }
\label{sum}
\end{center}
\end{figure}

\section{Conclusion and Outlook}
In general, SFOPT can occurs at different energy scale in different NP model from different motivation. 
The schematic phase transition GW signals are shown in Fig.~\ref{sum}.
GW becomes a new and realistic approach to explore DM,  baryogenesis and other NP since there are more and more relevant experiments, aLIGO, LISA, SKA, FAST, Tianqin, Taiji)\cite{Huang:2017laj,Yu:2016tar,Huang:2016odd,Huang:2018aja,Huang:2018lxq}.  

\section{Acknowlegments}
This work is supported by IBS under the project code IBS-R018-D1.

\end{document}